\documentclass{elsart}

 \usepackage{graphics}
 \usepackage{graphicx}
\usepackage{amssymb}

\begin{document}

\begin{frontmatter}


% use the thanksref command within \title, \author or \address for footnotes:
% \author{Name\thanksref{label2}}
% \thanks[label3]{Via A.Valerio 2, I-341127 Trieste, Italy}
% including your email address:
% \address{Address\thanksref{email}}
% \thanks[email]{E-mail: paolo.milazzo@trieste.infn.it}

\title{
Contemporary presence of dynamical and statistical production 
of intermediate mass fragments in midperipheral $^{58}$Ni+$^{58}$Ni 
collisions at 30 MeV/nucleon
}

 \author[ts]{P.M.Milazzo},
 \author[bo]{G.Vannini},
 \author[ts]{M.Sisto},
 \author[lns]{C.Agodi},
 \author[lns]{R.Alba},
 \author[lns]{G.Bellia},
 \author[minn]{M.Belkacem},
 \author[bo]{M.Bruno},
 \author[lns]{M.Colonna},
 \author[ba]{N.Colonna},
 \author[lns]{R.Coniglione},
 \author[bo]{M.D'Agostino},
 \author[lns]{A.Del Zoppo},
 \author[mi]{L.Fabbietti},
 \author[lns]{P.Finocchiaro},
 \author[lnl]{F.Gramegna},
 \author[mi]{I.Iori},
 \author[lns]{K.Loukachine},
 \author[lns]{C.Maiolino},
 \author[ts]{G.V.Margagliotti},
 \author[lnl]{P.F.Mastinu},
 \author[lns]{E.Migneco},
 \author[mi]{A.Moroni},
 \author[lns]{P.Piattelli},
 \author[ts]{R.Rui},
 \author[lns]{D.Santonocito},
 \author[lns]{P.Sapienza},
 \author[lns]{P.Ventura}

 \address[ts]{Dipartimento di Fisica and INFN, Trieste, Italy}
 \address[bo]{Dipartimento di Fisica and INFN, Bologna, Italy}
 \address[lns]{INFN, Laboratori Nazionali del Sud, Catania, Italy}
 \address[minn]{University of Minnesota, USA}
 \address[ba]{INFN, Bari, Italy}
 \address[mi]{Dipartimento di Fisica and INFN, Milano, Italy}
 \address[lnl]{INFN, Laboratori Nazionali di Legnaro, Italy}

\begin{abstract}
The $^{58}Ni+^{58}Ni$ reaction at 30 MeV/nucleon
has been experimentally investigated at the Superconducting Cyclotron
of the INFN Laboratori Nazionali del Sud.
In midperipheral collisions the production of massive fragments (4$\le$Z$\le$12),
consistent with the statistical fragmentation of the projectile-like residue
and the dynamical formation of a neck,
joining projectile-like and target-like residues, has been observed.
The fragments coming from these different processes 
differ both in charge distribution and isotopic composition. 
In particular it is shown that these mechanisms leading to fragment
production act contemporarily inside the same event.
\end{abstract}

\end{frontmatter}

% main text
%\section{}
%\label{}
%\begin{multicols}{2}
%\twocolumn
The production of intermediate mass fragments (IMF, 3$\le$Z$\le$20)
is the distinguishing feature of intermediate energy
heavy ion collisions.
A possible scenario for IMF emission involves the development of bulk
instabilities in the nuclear matter, which lead to a statistical
fragmentation. This is supported by
several experimental results concerning both
the decay of unique sources formed in central collisions
and the disassembly of quasitarget (QT) and quasiprojectile (QP) 
in dissipative peripheral collisions
\cite{tf,prcts}.

On the other hand theoretical studies of the collision dynamics predict the 
formation of highly deformed nuclei in this energy regime \cite{theo} and
experimental evidences have been collected on the role of the
formation and decay of neck-like structures in IMF production
\cite{exp-neck,Montoya}. Nevertheless it is still unclear
when this reaction mechanism sets in. For medium mass systems
Boltzmann-Nordheim-Vlasov (BNV) calculations predict the onset of neck
instabilities
to occur for energies between 30 and 70 MeV/nucleon \cite{theo}.
For lower energies the
interaction time between the projectile and the target is sufficiently
long for the neck to be reabsorbed by the QP and/or the QT, whereas
for higher energies the system evolves towards a fireball regime.

The study of the formation and decay of neck-like structures has a variety 
of theoretical implications. The beam energy at which a neck zone is
formed and decays depends on the Equation of State (EOS) of the
nuclear matter \cite{Sobotka}; a soft EOS favours neck
formation (because of a smaller orbiting) and rupture. Similarly an
isospin dependence of the EOS is also expected to influence the process
of formation and decay of an intermediate source and this should reflect
on the neutron-to-proton ratio of the emitted fragments \cite{Dempsey}.

The present analysis of the Ni+Ni midperipheral collisions is aimed to
investigate the interplay, inside the same event, between dynamically driven
neck instabilities and
QP statistical fragmentation in IMF production.

The experiment was performed at the INFN Laboratori Nazionali del Sud
with MEDEA \cite{Medea} and MULTICS \cite{Multics} apparata.
A beam of $^{58}$Ni at 30 MeV/nucleon bombarded a 2 mg/cm$^2$
thick nickel target.
The angular range 3$^{\circ}<\theta_{lab}<$28$^{\circ}$
was covered by the MULTICS array \cite{Multics}, consisting of 55 telescopes,
each of which was composed of an Ionization Chamber (IC), a Silicon
detector (Si) and a CsI crystal.
Typical energy resolutions
were 2\%, 1\% and 5\% for IC, Si and CsI, respectively.
The threshold for charge identification in the MULTICS array was about 1.5
MeV/nucleon. 
A good mass resolution for Z=1-6 isotopes was obtained
above 8.5, 10.5, 14 MeV/nucleon
for $^4$He, $^6$Li and $^{12}$C nuclei, respectively \cite{prcts}.
Light charged particles (Z=1,2) and $\gamma$-rays 
were detected at 30$^{\circ}<\theta_{lab}<$170$^{\circ}$ by
the BaF$_2$ ball of the MEDEA apparatus.
The geometric acceptance of the combined array was greater than 90\%
of 4$\pi$.

Since the relatively small number of detected fragments and particles
(Nc$\leq$5-6) makes not possible an accurate impact parameter selection,
using the standard methods
\cite{nc}, we followed a different approach: 1) selecting only the
"complete" multifragment events, i.e. events where at least three IMF were
produced and at least 80\% of the total linear momentum was detected;
2) defining peripheral and midperipheral collisions those for which the
heaviest fragment
(with charge at least 1/3 of that of the projectile, i.e. Z$\geq$9)
travels, in the laboratory frame, with a velocity higher than 80\% of
that of the projectile (v$_P$=7.6 cm/ns). Accordingly, since the energy
thresholds make not
possible the detection of the QT reaction products, we find
that the total detected charge (Z$_{TOT}$) does not differ from
that of the projectile for more than 30\% (20$\leq$Z$_{TOT}\leq$36).
Considering that not completely detected central events can simulate
complete non-central ones, we checked the existence in central events
of heavy fragments (Z$\geq$9) moving faster than 5.6 cm/ns, i.e. compatible
with the QP velocity. Central collision events have been selected
requiring that the heaviest fragment has a velocity close to the center
of mass one and that the total detected charge is at least 80\% of
the initial (projectile + target) value. We found that such a kind of
contamination is completely absent.

To get information about the impact parameters range
selected by this procedure and by the apparatus acceptance,
the experimental data have been compared
with the predictions of classical molecular dynamics (CMD) calculations 
\cite{CMD}.
CMD events are plotted in Fig.1a; the dot-filled area refers to the
amount of "complete" multifragmentation events
(with at least three IMF products) after experimental efficiency
filtering. In the dark area of Fig.1a
we present the filtered complete events with the further
condition that the heaviest fragment moves with a velocity higher than the
80\% of v$_P$. It becomes evident how midperipheral impact
parameter are preferentially selected.

\begin{figure}[htbp]
%\begin{figure}
\begin{center}
\includegraphics[width=10cm]{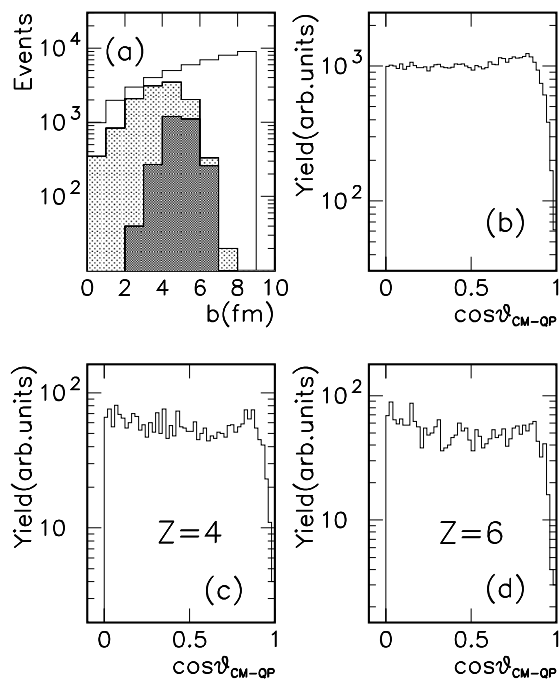}
\end{center}
\caption{(a)
Yields from CMD calculations: raw data (blank area), 
multifragmentation events after apparatus efficiency (dot-filled area), 
with further constraints on the heaviest fragments (dark area), (b-d) 
Angular distributions for IMF forward emitted by the quasiprojectile 
(all charged particle (b), Z=4 (c) and Z=6 (d) fragments). 
}
\end{figure}

The results presented hereafter
will refer only to midperipheral events with at least three IMF, with the aim 
of observing the IMF emitting sources. To this respect we present in Fig.2 the 
distributions of the parallel component of the velocity, with respect to the 
beam direction, for different Z values.
For carbon and oxygen nuclei two distinct distributions are evident: 
the first is centered at
6.5 cm/ns (the QP velocity) and the second one is at the center of mass
velocity (3.8 cm/ns), intermediate between that of the target and of
the projectile, because of the system symmetry.
Figg.2b-c show, in events where both the heaviest
fragment and an oxygen nucleus have velocities close to that of the
projectile, the yield of a third fragment (dot-dashed lines) .
It appears that this latter fragment has a great
probabilty of being emitted with a velocity centered around that of the c.m..
This is evidence that, while the QP is decaying in two or more fragments,
there is another system emitting IMF, intermediate (IS) between the
QP and QT.

\begin{figure}[htbp]
%\begin{figure}
\begin{center}
\includegraphics[width=10cm]{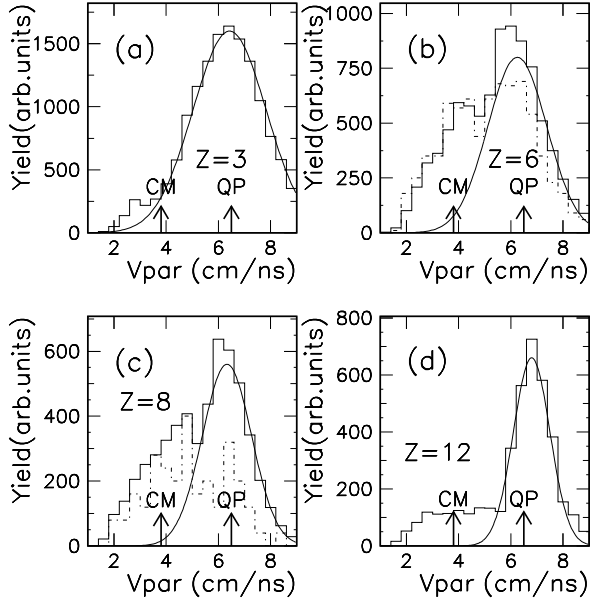}
\end{center}
\caption{
v$_{par}$ distributions for Z=3, 6 (dot-dashed line, multiplied by a factor 
10), 8 (dot-dashed line, multiplied by a factor 40), 12; the arrows refer to 
the center of mass (CM) and QP velocities.
}
\end{figure}

With the same event selection used for the fragment analysis for midperipheral
collisions we considered the proton spectra detected by the MEDEA apparatus.
Eight energy spectra, measured at different laboratory angles,
were simultaneously fitted
by the superposition of three Maxwellian distribution. The fit results put in
evidence the existence of a fast source (6.8 cm/ns and slope parameter
T$_{slope}\simeq$5.1 MeV) consistent with the QP, an almost at rest
source (QT system, 0.8 cm/ns, T$_{slope}\simeq$4.3 MeV) and an
intermediate velocity source (3.8 cm/ns) with high slope parameter
(T$_{slope}\simeq$10.1 MeV).
Moreover the QP and QT source show the same proton multiplicity. The
intermediate velocity component has been generally interpreted as also due to
pre-equilibrium emission in the interacting zone \cite{ct}.
These results are compatible with those shown in Ref.\cite{Rama}. 

These results show that in midperipheral collisions three
different emitting sources are present; there are events in which the IMF can
be simultaneously produced by the decay of QP and QT (its fragments are not seen
because under energy threshold for identification) sources and from
a neck, forming a midvelocity emission source.
Thus, disentangling the contributions from the QP and the neck sources
becomes a mandatory requirement to improve the understanding of the IMF
production mechanism and perform comparisons of the IMF experimental
yields with theoretical predictions.
To this purpose we first study the process leading to the disassembly of the
QP restricting the analysis to the fragments emitted
with v$_{par}>$6.5 cm/ns (see Fig.2). This constraint allows the selection of
the decay products forward emitted in the QP decay with negligible
contamination due to QT and midrapidity source emissions.
To check if the QP reaches an equilibration stage before its
de-excitation we studied in its reference frame the angular and energy 
distributions of the emitted fragments. 
Angular distributions are presented in Figg.1b-d; 
they are flat and then in agreement 
with the hypothesis of an isotropic emission; this is a 
necessary condition to establish a possible equilibration of the studied 
system. 
On the other hand fitting the energy distributions of different isotopes 
with Maxwellian functions
we get, for all the detected isotopes (3$\leq$A$\leq$14),
similar values (within errors) for the apparent temperatures $T_{slope}$.
This behaviour gives indications that the condition
of equilibration of the fragmenting systems is satisfied.
In Figg.3a-e the energy distributions of different isotopes are presented. 
Each panel (from a to d) refers to two different isotopes of
the same element to put in evidence the similarity of shapes and
slopes; in the (e) panel the Maxwellian fit is superimposed to the
experimental data showing the accuracy of the obtained results.

To gain more insight in the equilibration of the QP we compared the
experimental charge distribution with the microcanonical
Statistical Multifragmentation Model (SMM) predictions \cite{SMM},
performed for a Ni nucleus at one third of the normal density.
The events generated by SMM at different input excitation energies
were filtered by the experimental acceptance.
The experimental charge distribution is quite well reproduced
assuming an excitation energy of 4 MeV/nucleon (Fig.3f).

\begin{figure}[htbp]
%\begin{figure}
\begin{center}
\includegraphics[width=10cm]{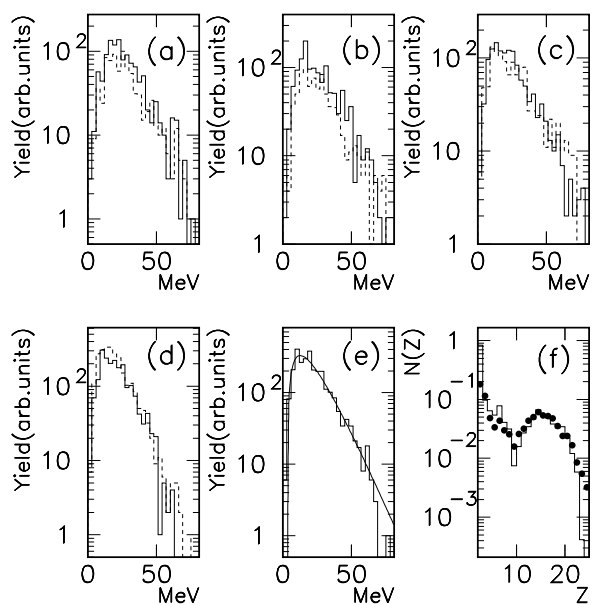}
\end{center}
\caption{
(a-e) Energy distributions for different
isotopes (a - $^{12}$C (full line), $^{13}$C (dashed line),
b - $^{10}$B (full), $^{11}$B (dashed), c - $^{7}$Be (full), $^{9}$Be (dashed),
d - $^{6}$Li (full), $^{8}$Li (dashed), e - Maxwellian fit for $^{7}$Li,
T$_{slope}$=8.9 MeV); 
(f) mean elemental event multiplicity N(Z) for QP fragments 
(solid point: experimental data, histogram: SMM predictions).
}
\end{figure}

The evidences of the equilibration of the QP source allow investigating some
characteristics, such as the temperature and the excitation energy.
We extract the temperature $T$ through the method of double
ratios of isotope yields \cite{albergo}.
As this method requires that the nuclei originate from the same emitting
source, only the Z$\geq$3 fragments forward emitted
(v$_{par}\geq$6.5 cm/ns, velocity of the QP in the laboratory frame)
were considered.
Table 1 gives the values for the most reliable thermometers
\cite{albergo,Betty}.
Since experimental temperature measurements are affected by secondary decays, 
we also report in Table 1 the values corrected as
suggested in Ref.\cite{Betty}.
Their mean value, $T_0$=3.9$\pm$0.3 MeV, can be considered as
the break-up temperature of the QP decaying system.

\begin{table}
\begin{center}
{\bf Table 1}
\end{center}
\caption{
Temperatures of the QP system extracted from different double yield isotope
ratio ($T_{exp}$) and calculated values after sequential feeding correction
($T_{corr}$).}
\begin{tabular}{lll}
\hline
      &$T_{exp}$(MeV)&$T_{corr}$(MeV) \\
\hline
$^6Li/^7Li-^{11}C/^{12}C$      &3.9$\pm$0.2  &3.3$\pm$0.3  \\
$^9Be/^{10}Be-^{11}C/^{12}C$   &6.7$\pm$0.9  &3.7$\pm$0.5  \\
$^{10}B/^{11}B-^{11}B/^{12}B$  &4.3$\pm$0.5  &4.3$\pm$0.5  \\
$^{11}B/^{12}B-^{11}C/^{12}C$  &4.3$\pm$0.3  &4.4$\pm$0.3  \\
$^{11}C/^{12}C-^{12}C/^{13}C$  &3.7$\pm$0.2  &3.6$\pm$0.2  \\
\hline
\end{tabular}
\end{table}

Besides the indication given by the statistical model,
we can give a rough estimate of the upper limit of the
excitation energy using the energy conservation and assuming that on
average there is an equal sharing of excitation energy between
QP and QT.
Then in the center of mass frame we 
have: $ E^*_{QP}={1\over 2}(m_Pv^2_P-m_{QP}v^2_{QP})+E^*_{NECK}$, 
where $m_P$, $m_{QP}$, $v_P$, $v_{QP}$ are mass and velocity of 
projectile and QP, respectively. 
Neglecting the amount of energy transferred to the
neck source (admittedly small to avoid its complete vaporization) and taking 
into account the mass difference between the projectile and the QP, 
the maximum of excitation energy ranges from 3.7 (if no nucleon transfer to
a neck is assumed) up to 5.9 MeV/nucleon
(half nickel is lost in the reaction); if, for
instance, we require the formation of an oxygen nucleus in the center of
mass source we have accordingly an estimate of $\simeq$4.5 MeV/nucleon.
This rough estimation
completely neglects other dissipation processes, as pre-equilibrium
emission.

The temperature and excitation energy values extracted in the present case 
are quite similar to
those measured \cite{prcts} for the QP fragmentation in the
$^{197}Au+^{197}Au$  reaction at 35 MeV/nucleon (T$_0$=3.9$\pm$0.2 MeV,
an upper limit of the excitation energy $\simeq$4.5 MeV/nucleon and
a corresponding measured (and also predicted by SMM) value of 
$\simeq$4 MeV/nucleon).
In Ref.\cite{prcts} it has been found good agreement between the excitation 
energy values obtained following different methods (calorimetric 
\cite{calo}, comparison 
with SMM predictions, rough estimation of the upper limit), 
in the range (3-6 MeV/nucleon). 

In conclusion the present data analysis shows that the QP source has attained
thermal equilibrium and that fragmentation is its main de-excitation
process, well reproduced by a statistical approach.

As already shown, in coincidence with the statistical
fragmentation of the QP, we observed the emission of IMF from a IS.
In the following we will discuss in detail the properties of this IMF emission
at midrapidity.
The expected distribution of the fragment velocity v$_{par}$ from the
multifragmentation of the QP source is Gaussian.
Thus the QP region of the v$_{par}$ distribution has been fitted, 
for each fragment Z value, with a Gaussian
function, giving the experimental QP yield (Y$_{QP}$). The fit parameters 
were extracted taking into account only the v$_{par}\geq$6.5 cm/ns part. 
In Fig. 2 the results are 
plotted, together with the experimental yields. Similar behaviors have
been found for all atomic numbers in the range 3$\leq$Z$\leq$14. 
Due to the experimental energy threshold (for the QT source side) 
the presented spectra are not symmetric around the c.m. velocity. 
To avoid possible contamination in the midrapidity region we 
evaluated the yield (Y$_{NECK}$) from the IS source 
as twice the difference between the whole v$_{par}$
distribution for velocities higher than that of the c.m. and Y$_{QP}$. 
This was done to avoid distorsions due to efficiency effects and 
possible QT contaminations for the lowest velocities. 
We checked that in the considered Z range the
experimental inefficiency doesn't affect the above-mentioned procedure
\cite{Multics}. 

In Fig.4a the ratio between the relative yields (Y$_{NECK}$/Y$_{QP}$)
is presented as a function of the atomic number Z. We observe a
bell-like shape, peaked around Z=9. We notice the high probability of
IMF emission from the neck zone when the event IMF multiplicity is at 
least 3 \cite{exp-neck}. The presence of this maximum
could be an effect of the breakup geometry; in fact, a rough
consistency has been found with percolative
calculations that compare emissions from a cylindrical shape neck,
joining spherical QT and QP, and from the
QP itself \cite{Montoya,Percola}.
If on one side the QP disassembly is ruled by
statistical models after thermal equilibrium has been reached, on
the other side the neck emission exhibit quite different features that cannot
be reproduced making statistical equilibrium assumptions.
We have thus performed BNV calculations using different EOS
parameters \cite{theo}. We found that with a compressibility term $K$ of
200 MeV (soft EOS) 
there is an evident massive neck formation (after 200 fm/c), that
is not reabsorbed by the QP or the QT (this behaviour disappears
increasing the $K$ values).
These calculations predict that on average we have a Z=8
fragment in the neck zone, and show that the IS fragment production
comes from material which is ``surface-like''
(since it originates from the overlap of the surfaces of the two nuclei)
and which could be neutron rich \cite{Dempsey}.
To investigate the neutron content of the neck matter we observed 
the $^6He$ experimental yield. 
Its emission is quasi negligible from a statistical decay, both because
the binding energy favours, for Z=2, $\alpha$ particle emission and because 
the N/Z value of $^6He$ is quite different from the corresponding ratio of the 
system (N/Z=2 for $^6He$,N/Z=30/28$\simeq$1 for the system).
If the surfaces of the interacting nuclei are more neutron rich than the 
bulk matter or the dynamical process leading to the formation of a neck 
structure has a particular isospin dependence, 
the $^6He$ production should be more abundant in the
neck zone with respect to the QP zone.
In Fig. 4b the experimental $^4He$ and $^6He$ yields are plotted versus 
$v_{par}$: one can see that, while the $^4He$ distribution is centered 
around the QP velocity (with a small emission from the neck zone) the $^6He$ 
yield is very scarse in the QP zone and starts to increase going towards the 
midvelocity region. 
One should take into account that the energy thresholds 
for mass identification are greater than for Z identification and that in 
the $He$ case they produce cuts for velocities lower than 4 cm/ns. 
To better see the isospin effect in Figg.4c-d 
the Y($^6He$)/Y($^4He$) and Y($^3He$)/Y($^4He$) yield ratios are plotted. 
The amount of more rich (poor) neutron $He$-isotopes increases (decreases) 
going towards the midvelocity region. 
The behaviour of these ratios clearly shows the increase of neutron content in 
the midvelocity zone 
with respect to the interacting matter. This is a further indication that 
in midperipheral collisions we observe an IMF production which
is due to two different mechanisms: one of statistical and the other
of dynamical nature.

\begin{figure}[htbp]
%\begin{figure}
\begin{center}
\includegraphics[width=10cm]{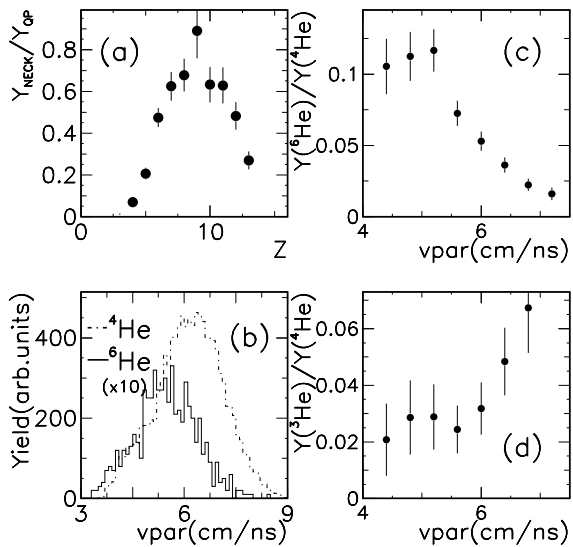}
\end{center}
\caption{
(a) Ratio of the measured yields for neck fragmentation and 
QP emission; (b) Yields of $^4$He (dot-dashed line) and $^6$He (full line, 
multiplied by a factor 10); Y($^6He$)/Y($^4He$) (c) and Y($^3He$)/Y($^4He$) (d)
yield ratios as a function of v$_{par}$.
}
\end{figure}

In conclusion, in the study of the Ni+Ni 30 MeV/nucleon dissipative
midperipheral collisions
it has been possible to reveal events in which the IMF's are emitted by 
two different sources with different mechanisms. 
We are in presence of a QP (and a QT),
with an excitation energy which leads to the multifragmentation regime; its
decay can be fully explained in terms of a statistical disassembly of
a thermalized system (T=3.9$\pm$0.2 MeV, E$^*\simeq$4 MeV/nucleon).
Contemporary to the IMF production
from the QP source an intermediate source is formed, emitting both
light particles and IMF. These fragments are more neutron rich than the
average matter of the overall system and have a quite different charge
distribution, with respect to the ones statistically emitted from the QP.
These features can be considered as a signature of the dynamical origin of
the midvelocity emission.
The presented results then show that IMF can be produced via different
mechanisms that can find contemporary room inside the same collision.

%\end{multicols}
\end{document}